\documentclass[12pt]{article}

\usepackage[a4paper,text={16.3cm,22cm}]{geometry}
\usepackage{amsmath,amsfonts,braket,amssymb,bm,bbm,xcolor,graphicx}
\usepackage[small,labelfont=bf]{caption}
\usepackage{cite}
\usepackage[
    bookmarksnumbered=true,
    urlcolor=blue,
    linkbordercolor=red,
    citebordercolor=green,
    bookmarksopen=true
    ]{hyperref}

\allowdisplaybreaks
\numberwithin{equation}{section}

\newcommand{\tr}[1]{\big\langle #1 \big\rangle}
\newcommand{\GGammac}{{\rm{I}}\hspace{-0.8mm}\Gamma^c}
\newcommand{\spac}{{\hspace{0.3mm}}}

\title{
Resummation of Glauber Phases in Non-Global
\texorpdfstring{\\[1ex]}{}
LHC Observables for Large \texorpdfstring{$N_c$}{Nc}
}
  
\begin{document}

\begin{titlepage}

\begin{flushright}
{\small
MITP-24-053\\
July 1, 2024
}
\end{flushright}

\makeatletter
\vskip0.8cm
\pdfbookmark[0]{\@title}{title}
\begin{center}
{\Large \bf\boldmath \@title}
\end{center}
\makeatother

\vspace{0.5cm}
\begin{center}
    \textsc{Philipp Böer,$^a$ Patrick Hager,$^a$ Matthias Neubert,$^{a,b}$ \\ Michel Stillger$^a$ and Xiaofeng Xu$^a$} \\[6mm]
    
    \textsl{${}^a$PRISMA$^+$ Cluster of Excellence \& Mainz Institute for Theoretical Physics\\
    Johannes Gutenberg University, Staudingerweg 9, 55128 Mainz, Germany\\[0.3cm]
    ${}^b$Department of Physics \& LEPP, Cornell University, Ithaca, NY 14853, U.S.A.}
\end{center}

\vspace{0.6cm}
\pdfbookmark[1]{Abstract}{abstract}
\begin{abstract}
The Glauber series for non-global jet observables at hadron colliders simultaneously includes the super-leading logarithms alongside an arbitrary number of Glauber phases. Building on the formalism of~\cite{Boer:2024hzh}, it is shown that the leading terms in this series for large $N_c$ can be resummed in closed form in renormalization-group improved perturbation theory. This remarkable observation suggests that large-$N_c$ methods might also be helpful to study other aspects of non-global logarithms at hadron colliders, and to combine our analytic results with amplitude-level parton showers. 
\end{abstract}

\vfill\noindent\rule{0.4\columnwidth}{0.4pt}\\
\hspace*{2ex} {\small \textit{E-mail:} \href{mailto:pboeer@uni-mainz.de}{pboeer@uni-mainz.de}, \href{mailto:pahager@uni-mainz.de}{pahager@uni-mainz.de}, \href{mailto:matthias.neubert@uni-mainz.de}{matthias.neubert@uni-mainz.de}, \\
\hspace*{2ex} \phantom{E-mail: } \href{mailto:m.stillger@uni-mainz.de}{m.stillger@uni-mainz.de}, \href{mailto:xiaxu@uni-mainz.de}{xiaxu@uni-mainz.de}}

\end{titlepage}

\section{Introduction}

Precision studies of jet observables at high-energy hadron colliders play an important role in testing the Standard Model of particle physics, because jet processes closely resemble the underlying hard-scattering dynamics. However, jet cross sections are also very challenging to calculate theoretically. The distinction of the highly collimated energetic particles constituting the jets from the soft radiation into the gaps between jets by imposing veto criteria in certain phase-space regions makes these observables ``non-global''. Such vetoes introduce additional scales in the problem, which in turn give rise to large logarithmic corrections, whose structure can be very complicated. So-called ``non-global logarithms'' (NGLs) arise from soft-gluon radiation off secondary emissions inside the jets~\cite{Dasgupta:2001sh}. Due to their intricate pattern and the complexity of the color algebra involved, the higher-order structure of these logarithms is highly non-trivial. At $e^+ e^-$ colliders, the resummation of the leading NGLs in the large-$N_c$ limit was accomplished by solving a non-linear integro-differential equation derived in~\cite{Banfi:2002hw}. The resummation at finite $N_c$ has been studied in~\cite{Weigert:2003mm,Hatta:2013iba,Hatta:2020wre,DeAngelis:2020rvq,Becher:2015hka,Becher:2016mmh}, and resummations at next-to-leading logarithmic accuracy have been achieved in~\cite{Banfi:2021owj,Banfi:2021xzn,Becher:2021urs,Becher:2023vrh,FerrarioRavasio:2023kyg}.

Due to the breakdown of color coherence from soft Glauber-gluon exchanges between the initial-state partons in the hard scattering process, the theory of non-global observables at hadron colliders is yet more complicated~\cite{Catani:2011st,Forshaw:2012bi,Schwartz:2017nmr}. This effect gives rise to so-called ``super-leading logarithms'' (SLLs)~\cite{Forshaw:2006fk,Forshaw:2008cq,Keates:2009dn} -- a class of double-logarithmic corrections at higher orders of perturbation theory of the form $(i\pi)^2\spac\alpha_s^{n+3} L^{2n+3}$ with $L=\ln(Q/Q_0)$, where $Q$ is of order the partonic center-of-mass energy, while the low scale $Q_0$ is the characteristic scale of the jet veto. The resummation of SLLs for arbitrary partonic scattering processes has been accomplished in~\cite{Becher:2021zkk,Becher:2023mtx}. Treating the Glauber phase $(i\pi)$ resulting from the imaginary part of the large logarithm $\ln(-Q/Q_0)$ as a large parameter, one can generalize the SLLs to the ``Glauber series'' of logarithmically-enhanced corrections of the form $(i\pi)^{2\ell}\spac\alpha_s^{n+1+2\ell} L^{2n+1+2\ell}$, where $\ell\in\mathbbm{N}$ denotes the number of pairs of Glauber exchanges. The resummation of the terms with $\ell>1$ in terms of multiple infinite sums has been studied in~\cite{Boer:2023jsy,Boer:2023ljq}. 

In recent work~\cite{Boer:2024hzh}, we have devised a strategy to systematically treat the resummation of the terms in the Glauber series in renormalization-group (RG) improved perturbation theory, i.e.\ including the scale dependence of the running coupling $\alpha_s(\mu)$ in the solution of the evolution equations. In this formalism, each Glauber-gluon insertion leads to an integral over a scale variable, which needs to be performed numerically. The present work is a direct sequel to~\cite{Boer:2024hzh}, in which we extend this formalism and derive quasi-analytic expressions for the large-$N_c$ limit of the Glauber series. As this series is subleading in $N_c$, these results correspond to a resummation of terms suppressed by $1/N_c^2$ compared to the leading contribution of the respective cross sections. We find that in this limit the structure of the Glauber series simplifies significantly and allows for an all-order resummation in terms of an expression involving at most four integrals. 
The analytic expressions we obtain will help to support and validate ongoing efforts to develop amplitude-level parton showers including quantum interference effects, see e.g.~\cite{AngelesMartinez:2018cfz,Nagy:2019pjp,Forshaw:2019ver,Hoche:2020pxj,Forshaw:2020wrq,DeAngelis:2020rvq,Hamilton:2020rcu,vanBeekveld:2022zhl,Becher:2023vrh,FerrarioRavasio:2023kyg}. 

The key insight of~\cite{Boer:2024hzh} is that at leading-logarithmic accuracy, the Glauber series can be represented as
\begin{equation}\label{eq:sigmaSLLG}
   \sigma_{2\to M}^{\rm SLL+G}(Q_0) = \sum_\text{partonic channels} \int\!d\xi_1 \int\!d\xi_2\,f_1(\xi_1,\mu_s)\,f_2(\xi_2,\mu_s) \, \hat{\sigma}_{2\to M}^{\rm SLL+G}(\xi_1,\xi_2,\mu_s) \,,
\end{equation}
where
\begin{equation}\label{eq:sigmaSLLGpartonic}
    \hat{\sigma}_{2\to M}^{\rm SLL+G}(\xi_1,\xi_2,\mu_s) = \sum_{l=1}^\infty \tr{\bm{\mathcal{H}}_{2\to M}(\xi_1,\xi_2,\mu_h)\,\bm{X}^T} \, \mathbbm{U}_{\rm SLL}^{(l)}(\mu_h,\mu_s)\,\varsigma
\end{equation}
is the partonic $2\to M$ scattering cross section, $f_i(\xi_i,\mu_s)$ denote the parton distribution functions, and $\bm{\mathcal{H}}_{2\to M}(\xi_1,\xi_2,\mu_h)$ are the hard functions in the factorization formula for $2\to M$ jet processes~\cite{Balsiger:2018ezi,Becher:2021zkk,Becher:2023mtx}. These functions are evaluated at the soft and hard scales, $\mu_s \sim Q_0$ and $\mu_h \sim \sqrt{\hat{s}}$, where $\sqrt{\hat{s}}$ is the partonic center-of-mass energy. Moreover, $\bm{X}$ denotes a vector containing the basis structures in color space derived in~\cite{Boer:2023jsy,Boer:2023ljq}, see e.g.~\eqref{eq:Xbasis} for quark-initiated processes, and $\varsigma^T=(1,0,\dots,0)$ with a single non-zero entry. The matrix $\mathbbm{U}_{\rm SLL}^{(l)}(\mu_h,\mu_s)$ is the corresponding representation of the RG-evolution operator, which evolves the hard function from $\mu_h$ to $\mu_s$, thereby resumming large logarithms $L_s=\ln(\mu_h/\mu_s)$ in the scale ratio alongside $l$ Glauber phases. It is given by 
\begin{equation}\label{eq:masterints}
\begin{aligned}
    \mathbbm{U}_{\rm SLL}^{(l)}(\mu_h,\mu_s)
    &= ( i\pi)^l N_c^{l-1}\,\frac{2^{l+3}}{\beta_0^{l+1}} 
    \int_1^{x_s}\!\frac{dx_l}{x_l}\,\ln\frac{x_s}{x_l} \int_1^{x_l}\!\frac{dx_{l-1}}{x_{l-1}}\,\ldots \int_1^{x_2}\!\frac{dx_1}{x_1}
    \\
    &\quad\times \mathbbm{U}_c(\mu_h,\mu_1) \Bigg[ \prod_{i=1}^{l-1}\,\mathbbm{V}^G \,\mathbbm{U}_c(\mu_i,\mu_{i+1}) \Bigg] \,,
\end{aligned}
\end{equation}
where $x_i=\alpha_s(\mu_i)/\alpha_s(\mu_h)$, and 
\begin{equation}\label{eq:Ucexp}
   \mathbbm{U}_c(\mu_i,\mu_j) = \exp\biggl[ N_c\,\GGammac\!\int_{\mu_j}^{\mu_i}\!\frac{d\mu}{\mu}\,\gamma_{\rm cusp}\big(\alpha_s(\mu)\big)
    \ln\frac{\mu^2}{\mu_h^2} \biggr]
\end{equation}
are matrix-valued Sudakov operators. In color and multiplicity space, the evolution equation for the hard function contains a logarithmically-enhanced soft-collinear anomalous dimension $\bm{\Gamma}^c$ and the Glauber operator $\bm{V}^G$~\cite{Becher:2021zkk}, whose matrix representations in the color basis are denoted by $\GGammac$ and $\mathbbm{V}^G$, respectively. As is evident from~\eqref{eq:masterints}, $l$ counts the number of Glauber-operator insertions. After diagonalization of the matrices $\GGammac$, the matrix exponential~\eqref{eq:Ucexp} can be expressed through the scalar functions 
\begin{equation} \label{eq:scalar_evolution_function}
   U_c(v; \mu_i,\mu_j) = \exp\biggl[ v\spac N_c \int_{\mu_j}^{\mu_i}\!\frac{d\mu}{\mu}\,\gamma_{\rm cusp}\big(\alpha_s(\mu)\big) \ln\frac{\mu^2}{\mu_h^2} \biggr]\,,
\end{equation}
where $v$ denotes one of the eigenvalues of $\GGammac$.

\section{Quark-initiated processes}
\label{sec:quarks}

We begin by presenting the resummation of the Glauber series in the large-$N_c$ limit for quark-initiated scattering processes. In this case, the associated color basis consists of the five elements
\begin{equation}\label{eq:Xbasis}
\begin{aligned}
    \bm{X}_1 &= \sum_{j>2} J_j\,if^{abc}\,\bm{T}_1^a\spac\bm{T}_2^b\spac\bm{T}_j^c \,,
    &\qquad
    \bm{X}_4 &= \frac{1}{N_c}\,J_{12}\,\bm{T}_1\cdot\bm{T}_2 \,, \\
    \bm{X}_2 &= \sum_{j>2} J_j\,(\sigma_1-\sigma_2)\,d^{abc}\,\bm{T}_1^a\spac\bm{T}_2^b\spac\bm{T}_j^c \,,
    &\qquad
    \bm{X}_5 &= J_{12}\,\bm{1} \,, \\
    \bm{X}_3 &= \frac{1}{N_c} \sum_{j>2} J_j \, (\bm{T}_1 - \bm{T}_2)\cdot \bm{T}_j \,,
\end{aligned}
\end{equation}
which contain the angular integrals
\begin{equation}\label{eq:Jints}
\begin{aligned}
   J_j &= \int\frac{d\Omega(n_k)}{4\pi} \, \big(W_{1j}^k - W_{2j}^k\big) \, \Theta_{\rm veto}(n_k) \,, \\
   J_{12} &= \int\frac{d\Omega(n_k)}{4\pi}\,W_{12}^k\,\Theta_{\rm veto}(n_k) \,.
\end{aligned}
\end{equation}
The soft dipole is defined as
\begin{equation}
    W_{ij}^k = \frac{n_i\cdot n_j}{n_i\cdot n_k\,n_j\cdot n_k} \,,
\end{equation}
with light-like vectors $n_i=p_i/E_i$ for each parton. The vector $n_k$ is restricted to the phase-space region where the jet veto is applied. In addition, $\sigma_i=-1$ ($\sigma_i=1$) if parton $i$ is an (anti)-quark. The factors $N_c$ are chosen such that the contribution of each color structure is at most of $\mathcal{O}(1)$ in the large-$N_c$ limit. The matrix representations $\mathbbm{V}^G$ and $\mathbbm{U}_c(\mu_i,\mu_j)$ are given in~\cite{Boer:2023jsy,Boer:2024hzh}. For $\mathbbm{V}^G$ it reads 
\begin{align}\label{eq:GGammacVG}
    \mathbbm{V}^G &=
    \begin{pmatrix}
        0 & -2\spac\delta_{q\bar q}\,\frac{N_c^2-4}{N_c^2} & ~\frac{4}{N_c^2} & ~~0 & ~~0 \\
        - \frac12 & 0 & ~0 & ~~0 & ~~0 \\
        1 & 0 & ~0 & ~~0 & ~~0 \\
        0 & 0 & ~0 & ~~0 & ~~0 \\
        0 & 0 & ~0 & ~~0 & ~~0
    \end{pmatrix} ,
\end{align}
where $\delta_{q\bar q}=\frac14\spac(\sigma_1-\sigma_2)^2$ equals~1 for the $q\bar q'$ initial states, and~0 for $q q'$ or $\bar q\bar q'$ initial states. The matrix exponential $\mathbbm{U}_c(\mu_i,\mu_j)$ in~\eqref{eq:Ucexp} takes the form
\begin{equation}\label{eq:Ucdiag}
    \mathbbm{U}_c(\mu_i,\mu_j) =
    \begin{pmatrix}
        U_c(1; \mu_i,\mu_j) & 0 & 0 & 0 & ~0 \\
        0 & U_c(1; \mu_i,\mu_j) & 0 & 0 & ~0 \\
        0 & 0 & U_c(\textstyle{\frac12}; \mu_i,\mu_j) & 0 & ~0 \\
        0 & 0 & 2 \big[ U_c(\textstyle{\frac12}; \mu_i,\mu_j) - U_c(1; \mu_i,\mu_j) \big] & ~U_c(1; \mu_i,\mu_j)~ & ~0 \\
        0 & 0 & \frac{2\spac C_F}{N_c} \big[ 1 - U_c(\textstyle{\frac12}; \mu_i,\mu_j) \big] & 0 & ~1
    \end{pmatrix} .
\end{equation}
Using these expressions, the first four terms $\mathbbm{U}_{\rm SLL}^{(l)}(\mu_h,\mu_s)$ have been derived in~\cite{Boer:2024hzh}. For odd values of $l$, for example, one finds
\begin{equation}\label{eq:USLL13}
\begin{aligned}
    \mathbbm{U}_{\rm SLL}^{(1)}(\mu_h,\mu_s)\,\varsigma
    &= \frac{16i\pi}{\beta_0^2} \int_1^{x_s}\!\frac{dx_1}{x_1}\,\ln\frac{x_s}{x_1}\,
     U_c(1; \mu_h,\mu_1)\,\varsigma \,, \\
    \mathbbm{U}_{\rm SLL}^{(3)}(\mu_h,\mu_s)\,\varsigma 
    &= - \frac{64\spac i\pi^3}{\beta_0^4}\,N_c^2 \int_1^{x_s}\!\frac{dx_3}{x_3}\,\ln\frac{x_s}{x_3} \int_1^{x_3}\!\frac{dx_2}{x_2}
     \int_1^{x_2}\!\frac{dx_1}{x_1} \\
    &\quad\times\big[ K_{12}\,U_c(1;\mu_h,\mu_3) 
     + \textstyle{\frac{4}{N_c^2}}\,U_c(1,\textstyle{\frac12},1; \mu_h,\mu_1,\mu_2,\mu_3) \big] \varsigma \,,
\end{aligned}
\end{equation}
with
\begin{equation}
    K_{12} = \frac{N_c^2-4}{N_c^2}\,\delta_{q\bar q} = \delta_{q\bar q} + \mathcal{O}(1/N_c^2) \,,
\end{equation}
and we have used the short-hand notation
\begin{equation}\label{eq:UcChains}
   U_c(v^{(1)},\dots,v^{(l)}; \mu_h,\mu_1,\dots,\mu_l)
   \equiv U_c(v^{(1)}; \mu_h,\mu_1)\,U_c(v^{(2)}; \mu_1,\mu_2) \dots U_c(v^{(l)}; \mu_{l-1},\mu_l)
\end{equation}
for the product of evolution factors. In the large-$N_c$ limit, only the term proportional to $K_{12}=\delta_{q\bar q}$ in the last line of~\eqref{eq:USLL13} prevails. These terms are associated with significantly simpler evolution factors, independent of $x_1$ and $x_2$, such that these integrations can be easily performed. One finds
\begin{equation}
   \mathbbm{U}_{\rm SLL}^{(3)}(\mu_h,\mu_s)\,\varsigma 
   \to \frac{16 i\pi}{\beta_0^2}\,\int_1^{x_s}\!\frac{dx_3}{x_3}\,\ln\frac{x_s}{x_3}\,\delta_{q\bar q}\,
    \frac{1}{2!} \bigg(\frac{2 \spac i\pi N_c}{\beta_0}\,\ln x_3\bigg)^{\!2}\,U_c(1;\mu_h,\mu_3)\,\varsigma \,,
\end{equation}
which has a similar structure as $\mathbbm{U}_{\rm SLL}^{(1)}(\mu_h,\mu_s)$ up to the additional factor in the integrand. This feature persists for higher $l$, and, therefore, all higher-order contributions in the Glauber series are absent for $qq$ and $\bar q \bar q$ scattering.

For large $N_c$, the structure of $\mathbbm{U}_c(\mu_i,\mu_j)$ remains unchanged but the (1,3) entry of $\mathbbm{V}^G$~\eqref{eq:GGammacVG} vanishes, which leads to a substantial simplification. In this case, the relevant block in the Glauber series~\eqref{eq:masterints} fulfills
\begin{equation}\label{eq:largeNc_master}
   \mathbbm{V}^G\,\mathbbm{U}_c(\mu_i,\mu_j) = \mathbbm{V}^G \, U_c(1;\mu_i,\mu_j) \,,
\end{equation}
allowing us to rewrite
\begin{equation} \label{eq:coef_quarks}
\begin{aligned}
   \mathbbm{U}_{\rm SLL}^{(l)}(\mu_h,\mu_s) \, \varsigma
   &= (i\pi)^l N_c^{l-1}\,\frac{2^{l+3}}{\beta_0^{l+1}} \int_1^{x_s}\!\frac{dx_l}{x_l}\,\ln\frac{x_s}{x_l}
    \int_1^{x_l}\!\frac{dx_{l-1}}{x_{l-1}}\,\ldots \int_1^{x_2}\!\frac{dx_1}{x_1} \\
   &\quad\times \mathbbm{U}_c(\mu_h,\mu_1) \big(\mathbbm{V}^G \big)^{l-1} \, U_c(1;\mu_1,\mu_l) \, \varsigma \,.
\end{aligned}
\end{equation}
In the case of odd $l=2\ell-1$, the remaining matrix structure simplifies as $\varsigma$ is an eigenvector of $(\mathbbm{V}^G)^2$ with eigenvalue $\delta_{q\bar q}$. Exploiting further that $\mathbbm{U}_c(\mu_h,\mu_1)\,\varsigma=U_c(1;\mu_h,\mu_1)\,\varsigma$, one can trivially perform the $x_i$-integrals for $i<l$, yielding
\begin{equation}
\begin{aligned}
   \mathbbm{U}_{\rm SLL}^{(2\ell-1)}(\mu_h,\mu_s)\,\varsigma 
   &= 8\,\frac{(2i\pi)^{2\ell-1}}{\beta_0^{2\ell}} \, \frac{N_c^{2\ell-2}}{(2\ell-2)!}\,
    \big[ \delta_{1\ell} + \delta_{q\bar q} ( 1- \delta_{1\ell}) \big] \\
    &\quad\times \int_1^{x_s}\!\frac{dx_l}{x_l}\,\ln\frac{x_s}{x_l}\,\ln^{2\ell-2}x_l\,U_c(1; \mu_h,\mu_l) \, \varsigma \,.
\end{aligned}
\end{equation}
While for even $l=2\ell$ the vector $\varsigma$ is \emph{not} an eigenvector of $(\mathbbm{V}^G)^{2\ell-1}$, one can use
\begin{equation}
    \big(\mathbbm{V}^G\big)^{2\ell-1} \, \varsigma = \delta_{q\bar q}^{\ell-1} \, \mathbbm{V}^G \, \varsigma
\end{equation}
and  perform the integrals over the $x_i$ variables for $1<i<l$ to obtain
\begin{align} \label{eq:U2l}
    \mathbbm{U}_{\rm SLL}^{(2\ell)}(\mu_h,\mu_s)\,\varsigma
    &= 8\,\frac{(2i\pi)^{2\ell}}{\beta_0^{2\ell+1}}\,\frac{N_c^{2\ell-1}}{(2\ell-2)!}\,
   \big[ \delta_{1\ell} + \delta_{q\bar q} ( 1- \delta_{1\ell}) \big]
    \int_1^{x_s}\!\frac{dx_l}{x_l}\,\ln\frac{x_s}{x_l} \int_1^{x_l}\!\frac{dx_1}{x_1}\,\ln^{2\ell-2}\frac{x_l}{x_1} \nonumber\\
    &\quad\times
    \begin{pmatrix}
        0 \\
        - \frac12\,U_c(1;\mu_h,\mu_l) \\
        U_c(\frac12,1;\mu_h,\mu_1,\mu_l) \\
        2 \big[ U_c(\frac12,1;\mu_h,\mu_1,\mu_l) - U_c(1;\mu_h,\mu_l) \big] \\
        \frac{2\spac C_F}{N_c} \big[ U_c(1;\mu_1,\mu_l) - U_c(\frac12,1;\mu_h,\mu_1,\mu_l) \big]       
    \end{pmatrix} .
\end{align}
Note that while the factor $\frac{2\spac C_F}{N_c}=1-\frac{1}{N_c^2}$ in the last component of this expression should be replaced by~1 in the strict large-$N_c$ limit, we prefer to keep it in its original form, as this ensures that the terms with $l=2$, i.e.\ the SLLs, are reproduced exactly.

It is now straightforward to sum the Glauber series in the large-$N_c$ limit. For odd values of $l$, one finds
\begin{equation} \label{eq:resummed_odd_qq}
   \sum_{\ell=1}^\infty \mathbbm{U}_{\rm SLL}^{(2\ell-1)}(\mu_h,\mu_s)\,\varsigma 
   = \frac{16 i\pi}{\beta_0^2} \int_1^{x_s}\!\frac{dx_2}{x_2}\,\ln\frac{x_s}{x_2} \; U_c(1; \mu_h,\mu_2)
    \bigg[ 1 - 2\delta_{q\bar q}\,\sin^2\!\bigg( \frac{\pi N_c}{\beta_0}\,\ln x_2 \bigg) \bigg] \varsigma \,,
\end{equation}
where the right-hand side is proportional to the vector $\varsigma$, i.e.\ only its first component is non-zero. The sum over even values of $l$ yields a similar result with an additional non-trivial scale integral, as evident from~\eqref{eq:U2l}. The expression reads
\begin{align} \label{eq:resummed_even_qq}
    \sum_{\ell=1}^\infty \mathbbm{U}_{\rm SLL}^{(2\ell)}(\mu_h,\mu_s)\,\varsigma 
    &= \frac{16\pi}{\beta_0^2} \, \frac{2\pi N_c}{\beta_0} \int_1^{x_s}\!\frac{dx_2}{x_2}\,\ln\frac{x_s}{x_2}
     \int_1^{x_2}\!\frac{dx_1}{x_1} \bigg[ 1 - 2\delta_{q\bar q}\,\sin^2\!\bigg( \frac{\pi N_c}{\beta_0}\,\ln\frac{x_2}{x_1} \bigg) \bigg]
    \nonumber\\*
    &\quad\times
    \begin{pmatrix}
        0 \\
        \frac12\,U_c(1;\mu_h,\mu_2) \\
        -U_c(\frac12,1;\mu_h,\mu_1,\mu_2) \\
        2 \big[ U_c(1;\mu_h,\mu_2) - U_c(\frac12,1;\mu_h,\mu_1,\mu_2) \big] \\
        \frac{2\spac C_F}{N_c} \big[ U_c(\frac12,1;\mu_h,\mu_1,\mu_2) - U_c(1;\mu_1,\mu_2) \big]       
    \end{pmatrix} .
\end{align}
Combining~\eqref{eq:resummed_odd_qq} and~\eqref{eq:resummed_even_qq} yields the resummed Glauber series in RG-improved perturbation theory, containing at most two scale integrals. In these expressions, the 1 inside the square brackets corresponds to the SLLs, whereas the term proportional to $\delta_{q\bar q}$ accounts for the effects of higher-order Glauber phases. It is therefore evident that the latter effects always reduce the contributions of the SLLs, albeit typically by a small amount, see Section~\ref{sec:numerics}. Note that for QCD processes with tree-level hard functions, only an even number of Glauber operator insertions contribute, as the cross section must be real-valued. In general, however, the full Glauber series, containing both even and odd $l$ values, is relevant if the hard function features complex phases with different color structures, e.g.~for cross sections involving electroweak gauge bosons~\cite{Forshaw:2021fxs}.

\subsubsection*{Fixed-coupling results and asymptotic behavior}

To determine the asymptotic behavior of the resummed Glauber series in the limit where $\alpha_s \spac L_s^2 \gg 1$, one can work in the approximation of a fixed coupling $\alpha_s\equiv\alpha_s(\bar\mu)$, with $\bar{\mu}\in(\mu_s,\mu_h)$, as the running is a single-logarithmic effect. In this case, one can use 
\begin{equation} \label{eq:fixed_coupling_approx}
   x_i = \frac{\alpha_s(\mu_i)}{\alpha_s(\mu_h)} \approx 1 + \frac{\beta_0\spac\alpha_s}{2\pi}\,L_i \,,
\end{equation}
where $L_i=\ln(\mu_h/\mu_i)$, and derive fixed-coupling versions of~\eqref{eq:resummed_odd_qq} and~\eqref{eq:resummed_even_qq}. Using also that~\eqref{eq:scalar_evolution_function} evaluates in this case to 
\begin{equation}\label{eq:Sudakov}
   U_c(v; \mu_i,\mu_j) \equiv \exp\!\big[ v\,w \big( z_i^2 - z_j^2 \big) \big] ,
\end{equation}
where $L_i\equiv z_i\spac L_s$, $w=\frac{N_c\alpha_s(\bar\mu)}{\pi}L_s^2$, it is possible to determine the asymptotic behavior by applying the same techniques as in~\cite{Boer:2024hzh}. For odd $l$, one finds after rescaling $y_i=\sqrt{w}\spac z_i$ and introducing $w_\pi = \frac{N_c\alpha_s(\bar\mu)}{\pi}\pi^2$ 
\begin{align} \label{eq:resummed_fixed_coupling_qq_odd}
    \sum_{\ell=1}^\infty \mathbbm{U}_{\rm SLL}^{(2\ell-1)}(\mu_h,\mu_s)\,\varsigma 
    &= \frac{4i\spac\alpha_s\spac L_s}{\pi\spac N_c} \sqrt{w_\pi} \int_0^{\sqrt{w}}\!dy_2\,\Big(1-\frac{y_2}{\sqrt{w}}\Big) \, 
     e^{-y_2^2} \biggl[ 1 - 2\delta_{q\bar q}\,\sin^2\biggl( \frac{\sqrt{w_\pi}}{2}\,y_2 \biggr) \biggr] \varsigma \nonumber\\
    &= \frac{2i\spac\alpha_s\spac L_s}{\pi\spac N_c} \,\sqrt{w_\pi} 
     \bigg\{ \sqrt{\pi}\big[1 - \delta_{q\bar q} \big(1-e^{-\frac{w_\pi}{4}}\big)\big] \\ \nonumber
    &\hspace{30mm} - \frac{1}{\sqrt{w}} \bigg[1 - \delta_{q\bar q} \, \sqrt{w_\pi} \, F\bigg(\frac{\sqrt{w_\pi}}{2}\bigg)\bigg] 
     + \mathcal{O}(e^{-w}) \bigg\} \,\varsigma \,,
\end{align}
where the second relation is valid in the limit $w \gg 1$. The Dawson function is defined as 
\begin{equation}
   F(z) = e^{-z^2} \int_0^z\!dx\,e^{x^2} = \frac{\sqrt{\pi}}{2} \, e^{-z^2} \, \text{erfi}(z) \,.
\end{equation}
To obtain the asymptotic result, it suffices to replace the upper limit of the $y_2$ integral by infinity. For even $l$, the situation is more complicated, and one finds 
\begin{equation} \label{eq:resummed_fixed_coupling_qq_even}
\begin{aligned}
    \sum_{\ell=1}^\infty \mathbbm{U}_{\rm SLL}^{(2\ell)}(\mu_h,\mu_s)\,\varsigma 
    &= - \frac{4\alpha_s\spac L_s}{\pi\spac N_c}\,w_\pi
     \int_0^{\sqrt{w}}\!dy_2\,\Big(1-\frac{y_2}{\sqrt{w}}\Big) \, e^{-y_2^2} \int_0^{y_2}\!dy_1
     \\[-2mm]
    &\quad\times \biggl[ 1 - 2\delta_{q\bar q}\,\sin^2\biggl( \frac{\sqrt{w_\pi}}{2}\,(y_2-y_1) \biggr) \biggr]  
    \begin{pmatrix}
        0 \\
        - \frac12 \\
        e^{\frac{1}{2}\spac y_1^2} \\
        2\spac \big( e^{\frac{1}{2}\spac y_1^2} - 1 \big) \\
        \frac{2\spac C_F}{N_c} \big( e^{y_1^2} - e^{\frac{1}{2}\spac y_1^2} \big)
    \end{pmatrix} .
\end{aligned}
\end{equation}
In most cases it is sufficient to replace the upper integration boundary of the $y_2$ integral with $\sqrt{w} \to \infty$ to obtain the asymptotic form. The only exception is the fifth component, which due to the factor $e^{y_1^2}$ develops a logarithmic dependence on $w$.
Integrating up to infinity gives a divergent result, but the divergence is cancelled by a contribution from the region $y_2 \sim \sqrt{w}$.
One needs the expansions of the three integrals
\begin{align}
\begin{split} \label{eq:asymptotics_qq_even_1}
    &\int_0^{\sqrt{w}}\!dy_2 \, \Big(1-\frac{y_2}{\sqrt{w}}\Big) \, e^{-y_2^2} \int_0^{y_2}\!dy_1 \spac \bigg[ 1 - 2\delta_{q\bar q}\,\sin^2\!\bigg( \frac{\sqrt{w_\pi}}{2}\,(y_2-y_1) \bigg) \bigg] 
    \\
    &= \frac12 - \frac{\sqrt{\pi}}{4\sqrt{w}} - \delta_{q\bar q} \bigg[\frac12 - \frac{1}{\sqrt{w_\pi}} \, F\bigg(\frac{\sqrt{w_\pi}}{2}\bigg) - \frac{\sqrt{\pi}}{4\sqrt{w}} \Big(1 - e^{-\frac{w_\pi}{4}}\Big)\bigg] + \mathcal{O}(e^{-w}) \,,
\end{split}
\\[7mm]
\begin{split} \label{eq:asymptotics_qq_even_2}
    &\int_0^{\sqrt{w}}\!dy_2\,\Big(1-\frac{y_2}{\sqrt{w}}\Big) \, e^{-y_2^2} \int_0^{y_2}\!dy_1
    \biggl[ 1 - 2\delta_{q\bar q}\,\sin^2\biggl( \frac{\sqrt{w_\pi}}{2}\,(y_2-y_1) \biggr) \biggr] \, e^{\frac{1}{2}\spac y_1^2}
    \\
    &= \frac{\ln(1+\sqrt{2})}{\sqrt{2}} - \frac{\sqrt{\pi}}{2 \sqrt{2 w}} - \delta_{q\bar q} \biggl[ \frac{\ln(1+\sqrt{2})}{\sqrt{2}} + i \pi \sqrt{2} \, e^{\frac{w_\pi}{4}} \, T\bigg(\!\sqrt{w_\pi},\frac{i}{\sqrt2}\bigg)
    \\
    &\quad+ \frac{\pi \sqrt{w_\pi}}{4\sqrt{2w}} \, e^{\frac{w_\pi}{4}} \bigg(
    \text{erf}\spac\bigg(\frac{\sqrt{w_\pi}}{2}\bigg)
    - \text{erf}\spac\bigg(\frac{\sqrt{w_\pi}}{\sqrt{2}} \bigg) \bigg) \biggr] + \mathcal{O}(e^{-w/2}) \,, 
\end{split}
\intertext{and}
\begin{split} \label{eq:asymptotics_qq_even_3}
    &\int_0^{\sqrt{w}}\!dy_2\,\Big(1-\frac{y_2}{\sqrt{w}}\Big) \, e^{-y_2^2} \int_0^{y_2}\!dy_1
    \biggl[ 1 - 2\delta_{q\bar q}\,\sin^2\biggl( \frac{\sqrt{w_\pi}}{2}\,(y_2-y_1) \biggr) \biggr] \, e^{y_1^2}
    \\
    &= \frac{\ln(4w) + \gamma_E - 2}{4} - \delta_{q\bar q} \biggl[ \frac{w_\pi}{8} \, _2F_2\big(1,1;{\textstyle\frac32},2;-\frac{w_\pi}{4}\big) - \frac{\pi \sqrt{w_\pi}}{8\sqrt{w}} \, \text{erf}\bigg(\frac{\sqrt{w_\pi}}{2}\bigg) \biggr] + \mathcal{O}(w^{-1}) \,,
\end{split}
\end{align}
where 
\begin{equation}
   \text{erf}(z) = \frac{2}{\sqrt\pi} \int_0^z\!dx\,e^{-x^2}
\end{equation}
is the error function, ${}_2F_2(a_1,a_2;b_1,b_2;z)$ is a generalized hypergeometric function, and 
\begin{equation}
   T(x,a) = \frac{1}{2\pi} \int_0^a\!dt\,\frac{e^{-\frac12 x^2\spac(1+t^2)}}{1+t^2}
\end{equation}
denotes the Owen $T$-function. In agreement with the findings of~\cite{Boer:2024hzh}, it turns out that the dependence on the variable $w$ of the resummed Glauber series cancels in all but the fifth component, which has a residual logarithmic dependence on $w$.

\section{Gluons in the initial state}
\label{sec:gluons}

There are two reasons why the resummation of the Glauber series for quark-initiated processes is particularly simple in the large-$N_c$ limit.
First, the relevant building block $\mathbbm{V}^G\,\mathbbm{U}_c(\mu_i,\mu_j)$ fulfills relation~\eqref{eq:largeNc_master}, allowing us to perform all except (at most) two scale integrals and to simplify the matrix structure in~\eqref{eq:masterints}. 
Second, $\GGammac$ remains diagonalizable in this limit. As soon as gluons are present in the initial state, these properties are lost, complicating the resummation.

The color bases for quark-gluon and gluon-initiated processes, containing 14 and 20 elements, respectively, have been constructed in~\cite{Boer:2023ljq}. Rescaled with appropriate factors of $N_c$, see~\cite{Boer:2024hzh}, the matrices $\GGammac$ and $\mathbbm{V}^G$ are of $\mathcal{O}(1)$ in the large-$N_c$ expansion and can be found in Appendices~\ref{app:quark_gluon} and~\ref{app:gluon}. 
While in general the matrix $\GGammac$ has up to eleven different eigenvalues for processes with gluons in the initial state, in the large-$N_c$ limit these degenerate to at most five eigenvalues, given by
\begin{align} \label{eq:largeNc_eigenvalues}
    v \in \Big\{ 0 \,, \frac{1}{2} \,, 1 \,, \frac{3}{2} \,, 2 \Big\} \,,
\end{align}
where the eigenvalue $2$ only appears if both initial-state partons are gluons. Curiously, it turns out that in the large-$N_c$ limit all eigenvalues are half-integers ranging from zero up to the sum of the spins of the initial-state partons. 
It would be interesting to explore whether there is a deeper reason for this coincidence. 
As for quark-initiated processes, Sudakov factors with the eigenvalues $0$ and $\frac12$ only arise from the left-most insertion of $\mathbbm{U}_c(\mu_h,\mu_1)$ in~\eqref{eq:masterints}. Therefore, the generalization of~\eqref{eq:largeNc_master} reads\footnote{Note that $\mathbbm{V}^G_2$ depends on $\mu_i,\mu_j$. However, as explained below~\eqref{eq:Nc_dependent_eigenvalues}, this dependence is irrelevant.}
\begin{equation} \label{eq:largeNc_master_general}
    \mathbbm{V}^G\,\mathbbm{U}_c(\mu_i,\mu_j) 
    = \mathbbm{V}^G_1 \, U_c(1;\mu_i,\mu_j) + \mathbbm{V}^G_{3/2} \, U_c(\textstyle{\frac32};\mu_i,\mu_j)
     + \mathbbm{V}^G_2 \, U_c(2;\mu_i,\mu_j) \,.
\end{equation}

Since $\varsigma$ is an eigenvector of $\mathbbm{U}_c(\mu_i,\mu_j)$ with eigenvalue $U_c(1;\mu_i,\mu_j)$, irrespective of the nature of the initial-state partons, the coefficient matrices fulfill $\mathbbm{V}^G_{3/2}\,\varsigma=\mathbbm{V}^G_2\,\varsigma=0$. For quark-gluon-initiated processes $\mathbbm{V}^G_2=0$, because the eigenvalue $2$ does not appear. Furthermore, the coefficient matrices satisfy 
\begin{align}
    \mathbbm{V}^G_1\,\mathbbm{V}^G_{3/2} = 0 \,, \qquad 
    \mathbbm{V}^G_{3/2}\,\mathbbm{V}^G_2 = 0 \,, \qquad
    \mathbbm{V}^G_1\,\mathbbm{V}^G_2 = 0 \,, \qquad 
    \mathbbm{V}^G_2\,\mathbbm{V}^G_1 = 0 \,,
\end{align}
ensuring that only the three scalar functions $U_c(v,1;\mu_h,\mu_1,\mu_i)$, $U_c(v,\textstyle{\frac32},1;\mu_h,\mu_1,\mu_i,\mu_j)$ and $U_c(v,2,\textstyle{\frac32},1;\mu_h,\mu_1,\mu_i,\mu_j,\mu_k)$ contribute, where the first eigenvalue $v$ is arbitrary.

There is one more subtlety that needs to be taken into account when dealing with gluons in the initial state. Since $\GGammac$ is no longer diagonalizable once the large-$N_c$ limit is taken, its matrix exponential $\mathbbm{U}_c(\mu_i,\mu_j)$ with full $N_c$ dependence contains terms of the form 
\begin{equation} \label{eq:degenerate_eigenvalues_largeNc}
    N_c \big[U_c\big(\textstyle{\frac{3N_c+2}{2N_c}};\mu_i,\mu_j\big) - U_c\big(\textstyle{\frac{3N_c-2}{2N_c}};\mu_i,\mu_j\big)\big] 
    = 2\spac I_h(\mu_i,\mu_j) \, U_c(\textstyle{\frac32};\mu_i,\mu_j) + \mathcal{O}(1/N_c^2) \,,
\end{equation}
with the integral 
\begin{equation}
    I_h(\mu_i,\mu_j) \equiv N_c \int_{\mu_j}^{\mu_i}\!\frac{d\mu}{\mu}\,\gamma_{\rm cusp}\big(\alpha_s(\mu)\big) \ln\frac{\mu^2}{\mu_h^2} \,,
\end{equation}
which counts as $\mathcal{O}(1)$ in the large-$N_c$ expansion. This effect arises for all $N_c$-dependent eigenvalues of $\GGammac$, i.e.\
\begin{align} \label{eq:Nc_dependent_eigenvalues}
	v_{3,4} = \frac{3N_c\pm2}{2N_c} \,, \qquad 
	v_{5,6} = \frac{2(N_c\pm1)}{N_c} \,, \qquad 
	v_{9,10} = \frac{2N_c\pm1}{N_c}
\end{align}
in the notation of \cite{Boer:2024hzh}, where $v_3$, $v_5$, $v_9$ correspond to the plus signs. As a consequence, the coefficient matrix $\mathbbm{V}^G_2$ in~\eqref{eq:largeNc_master_general} depends on the scales $\mu_i$ and $\mu_j$. Fortunately, this dependence drops out in the only relevant product $\mathbbm{V}^G_2 \, \mathbbm{V}^G_{3/2}$. Therefore, we can treat $\mathbbm{V}^G_2$ as scale-independent in the following. The coefficient matrix $\mathbbm{V}^G_{3/2}$ does not depend on $\mu_i$ and $\mu_j$.

Inserting the decomposition~\eqref{eq:largeNc_master_general} in~\eqref{eq:masterints}, and exploiting the properties of the coefficient matrices, one finds 
\begin{equation} \label{eq:coef_gluons}
\begin{aligned}
    \mathbbm{U}_{\rm SLL}^{(l)}(\mu_h,\mu_s) \, \varsigma 
    &= (i\pi)^l N_c^{l-1}\,\frac{2^{l+3}}{\beta_0^{l+1}} \int_1^{x_s}\!\frac{dx_l}{x_l}\,\ln\frac{x_s}{x_l}
    \int_1^{x_l}\!\frac{dx_{l-1}}{x_{l-1}}\,\ldots \int_1^{x_2}\!\frac{dx_1}{x_1} \, \mathbbm{U}_c(\mu_h,\mu_1)
    \\
    &\quad\times \Big[ \big(\mathbbm{V}^G_1 \big)^{l-1} \, U_c(1;\mu_1,\mu_l) + \sum_{i=2}^{l-1} \big(\mathbbm{V}^G_{3/2} \big)^{i-1} 
    \big(\mathbbm{V}^G_1 \big)^{l-i} \, U_c(\textstyle{\frac32},1;\mu_1,\mu_i,\mu_l)
    \\
    &\qquad + \sum_{j=3}^{l-1} \sum_{i=2}^{j-1} \big(\mathbbm{V}^G_2 \big)^{i-1} \big(\mathbbm{V}^G_{3/2} \big)^{j-i} 
    \big(\mathbbm{V}^G_1 \big)^{l-j} \, U_c(2,\textstyle{\frac32},1;\mu_1,\mu_i,\mu_j,\mu_l) \Big] \, \varsigma \,.
\end{aligned}
\end{equation}
We start with the first term, which looks similar to the quark case, cf.~\eqref{eq:coef_quarks}. If $l\geq2$, one can trivially perform all integrals except the ones over $x_1$ and $x_l$. For $l=1$ there is only one such integral, and one can use that $\mathbbm{U}_c(\mu_h,\mu_1)\,\varsigma=U_c(1;\mu_h,\mu_1)\,\varsigma$. Combining these two cases, one obtains 
\begin{equation} \label{eq:scale_integrals_performed_1}
    \int_1^{x_l}\!\frac{dx_{l-1}}{x_{l-1}}\,\ldots \int_1^{x_2}\!\frac{dx_1}{x_1} = \delta_{1l} + (1-\delta_{1l}) \int_1^{x_l}\!\frac{dx_1}{x_1} \, \frac{1}{(l-2)!} \, \ln^{l-2} \frac{x_l}{x_1} \,.
\end{equation}
For the second and third term in~\eqref{eq:coef_gluons}, also the integrals over $x_i$ and $x_i,x_j$, respectively, are non-trivial. The remaining integrals evaluate to 
\begin{equation} \label{eq:scale_integrals_performed_2}
\begin{aligned}
    \int_1^{x_l}\!\frac{dx_{l-1}}{x_{l-1}}\,\ldots \int_1^{x_2}\!\frac{dx_1}{x_1} 
    &= \int_1^{x_l}\!\frac{dx_i}{x_i} \int_1^{x_i}\!\frac{dx_1}{x_1} \, \frac{1}{(i-2)!} \, \ln^{i-2}\frac{x_i}{x_1} \; 
     \frac{1}{(l-1-i)!} \, \ln^{l-1-i}\frac{x_l}{x_i} \,, \\[2mm]
    \int_1^{x_l}\!\frac{dx_{l-1}}{x_{l-1}}\,\ldots \int_1^{x_2}\!\frac{dx_1}{x_1} 
    &= \int_1^{x_l}\!\frac{dx_j}{x_j} \int_1^{x_j}\!\frac{dx_i}{x_i} \int_1^{x_i}\!\frac{dx_1}{x_1}\, 
     \frac{1}{(i-2)!}\,\ln^{i-2}\frac{x_i}{x_1} \\
    &\quad \times \frac{1}{(j-i-1)!} \, \ln^{j-i-1}\frac{x_j}{x_i} \; \frac{1}{(l-1-j)!} \, \ln^{l-1-j}\frac{x_l}{x_j} \,.
\end{aligned}
\end{equation}
Distinguishing the cases of even and odd $l$, one can calculate the matrix products in~\eqref{eq:coef_gluons} and perform the sums over $i,j$, and $l$, thereby accomplishing the resummation of the Glauber series.

For $l=2\ell-1$ odd, one finds for the matrix products 
\begin{equation}
\begin{aligned}
    \big(\mathbbm{V}^G_1 \big)^{2\ell-2} \, \varsigma &= \delta_{1\ell} \, \varsigma + (1-\delta_{1\ell}) \, \varsigma_1 \,,
    \\
    \big(\mathbbm{V}^G_{3/2} \big)^{i-1} \big(\mathbbm{V}^G_1 \big)^{2\ell-1-i} \, \varsigma &= (1+\delta_{i,2\ell-2}) \, \varsigma_{3/2} \,,
    \\
    \big(\mathbbm{V}^G_2 \big)^{i-1} \big(\mathbbm{V}^G_{3/2} \big)^{j-i} \big(\mathbbm{V}^G_1 \big)^{2\ell-1-j} \, \varsigma &= (1+\delta_{j,2\ell-2}) \, \varsigma_2 \,.
\end{aligned}
\end{equation}
The $\varsigma$ vectors differ between quark-gluon and gluon-initiated processes and are given in Appendices~\ref{app:quark_gluon} and~\ref{app:gluon}, respectively. Combining these results with~\eqref{eq:scale_integrals_performed_1} and~\eqref{eq:scale_integrals_performed_2}, the coefficient vector of the basis structures can be expressed as 
\begin{align} \label{eq:resummed_odd_general}
    \sum_{\ell=1}^\infty \mathbbm{U}_{\rm SLL}^{(2\ell-1)}(\mu_h,\mu_s)\,\varsigma 
    &= \frac{16i\pi}{\beta_0^2} \int_1^{x_s} \!\frac{dx_4}{x_4} \,\ln\frac{x_s}{x_4} \bigg\{ U_c(1;\mu_h,\mu_4) \, \varsigma
     \notag\\
    &\quad - \frac{2\pi N_c}{\beta_0} \int_1^{x_4} \!\frac{dx_1}{x_1} \, U_c(1;\mu_1,\mu_4) \, \sin\!\bigg(\frac{2\pi N_c}{\beta_0} \, 
     \ln \frac{x_4}{x_1}\bigg) \, \mathbbm{U}_c(\mu_h,\mu_1) \, \varsigma_1
     \notag\\
    &\quad - \bigg(\frac{2\pi N_c}{\beta_0}\bigg)^{\!2} \int_1^{x_4} \!\frac{dx_2}{x_2} \int_1^{x_2} \!\frac{dx_1}{x_1} \, 
     U_c(\textstyle{\frac32},1;\mu_1,\mu_2,\mu_4)
     \notag\\
    &\qquad \times \bigg[\cos\!\bigg(\frac{2\pi N_c}{\beta_0} \, \ln\frac{x_4}{x_1}\bigg) + \cos\!\bigg(\frac{2\pi N_c}{\beta_0} \, 
     \ln\frac{x_2}{x_1}\bigg) \bigg] \mathbbm{U}_c(\mu_h,\mu_1) \, \varsigma_{3/2}
     \notag\\
    &\quad + \bigg(\frac{2\pi N_c}{\beta_0}\bigg)^{\!3} \int_1^{x_4} \!\frac{dx_3}{x_3} \int_1^{x_3} \!\frac{dx_2}{x_2} 
     \int_1^{x_2} \!\frac{dx_1}{x_1} \, U_c(2,\textstyle{\frac32},1;\mu_1,\mu_2,\mu_3,\mu_4)
     \notag\\
    &\qquad \times \bigg[\sin\!\bigg(\frac{2\pi N_c}{\beta_0} \, \ln\frac{x_4}{x_1}\bigg) 
     + \sin\!\bigg(\frac{2\pi N_c}{\beta_0} \, \ln\frac{x_3}{x_1}\bigg) \bigg] \mathbbm{U}_c(\mu_h,\mu_1) \, \varsigma_2 \bigg\} .
\end{align}
After carrying out the last products $\mathbbm{U}_c(\mu_h,\mu_1)\,\varsigma_{\spac i}$, it is possible to combine some terms by integrating over $x_1$. However, these simplifications are not universal and are performed separately for quark-gluon and gluon-initiated processes in Appendices~\ref{app:quark_gluon} and~\ref{app:gluon}, respectively.

For $l=2\ell$ even, the matrix products evaluate to 
\begin{equation}
\begin{aligned}
    \big(\mathbbm{V}^G_1 \big)^{2\ell-1} \, \varsigma 
    &= \delta_{1\ell} \, \tilde\varsigma + (1-\delta_{1\ell}) \, \tilde\varsigma_1 \,, \\
    \big(\mathbbm{V}^G_{3/2} \big)^{i-1} \big(\mathbbm{V}^G_1 \big)^{2\ell-i} \, \varsigma 
    &= (1+\delta_{i,2\ell-1}) \, \tilde\varsigma_{3/2} \,, \\
    \big(\mathbbm{V}^G_2 \big)^{i-1} \big(\mathbbm{V}^G_{3/2} \big)^{j-i} \big(\mathbbm{V}^G_1 \big)^{2\ell-j} \, \varsigma 
    &= (1+\delta_{j,2\ell-1}) \, \tilde\varsigma_2 \,.
\end{aligned}
\end{equation}
The $\tilde\varsigma$ vectors are also given in Appendices~\ref{app:quark_gluon} and~\ref{app:gluon}. Performing the sums over $i,j$, and $\ell$ in~\eqref{eq:coef_gluons} results in 
\begin{equation} \label{eq:resummed_even_general}
\begin{aligned}
    \sum_{\ell=1}^\infty \mathbbm{U}_{\rm SLL}^{(2\ell)}(\mu_h,\mu_s)\,\varsigma 
    &= - \frac{16\pi}{\beta_0^2} \int_1^{x_s} \!\frac{dx_4}{x_4} \,\ln\frac{x_s}{x_4} 
    \bigg\{ \frac{2\pi N_c}{\beta_0} \int_1^{x_4} \!\frac{dx_1}{x_1}\,U_c(1;\mu_1,\mu_4) \, \mathbbm{U}_c(\mu_h,\mu_1) \, \tilde\varsigma
    \\
    &\quad - \frac{2\pi N_c}{\beta_0} \int_1^{x_4} \!\frac{dx_1}{x_1} \, U_c(1;\mu_1,\mu_4) \, 2\spac\sin^2\!\bigg(\frac{\pi N_c}{\beta_0} \,
    \ln \frac{x_4}{x_1}\bigg) \, \mathbbm{U}_c(\mu_h,\mu_1) \, \tilde\varsigma_1
    \\
    &\quad - \bigg(\frac{2\pi N_c}{\beta_0}\bigg)^{\!2} \int_1^{x_4} \!\frac{dx_2}{x_2} \int_1^{x_2} \!\frac{dx_1}{x_1} \,
    U_c(\textstyle{\frac32},1;\mu_1,\mu_2,\mu_4)
    \\
    &\qquad \times \bigg[\sin\!\bigg(\frac{2\pi N_c}{\beta_0} \, \ln\frac{x_4}{x_1}\bigg) 
    + \sin\!\bigg(\frac{2\pi N_c}{\beta_0} \, \ln\frac{x_2}{x_1}\bigg) \bigg] \mathbbm{U}_c(\mu_h,\mu_1) \, \tilde\varsigma_{3/2}
    \\
    &\quad - \bigg(\frac{2\pi N_c}{\beta_0}\bigg)^{\!3} \int_1^{x_4} \!\frac{dx_3}{x_3} \int_1^{x_3} \!\frac{dx_2}{x_2} 
    \int_1^{x_2} \!\frac{dx_1}{x_1} \, U_c(2,\textstyle{\frac32},1;\mu_1,\mu_2,\mu_3,\mu_4)
    \\
    &\qquad \times \bigg[\cos\!\bigg(\frac{2\pi N_c}{\beta_0} \, \ln\frac{x_4}{x_1}\bigg)
    + \cos\!\bigg(\frac{2\pi N_c}{\beta_0} \, \ln\frac{x_3}{x_1}\bigg) \bigg] \mathbbm{U}_c(\mu_h,\mu_1) \, \tilde\varsigma_2 \bigg\} .
\end{aligned}
\end{equation}
After the products $\mathbbm{U}_c(\mu_h,\mu_1)\,\tilde\varsigma_{\spac i}$ are performed, the result is very lengthy and contains terms proportional to $I_h(\mu_h,\mu_1)$ by the mechanism described in~\eqref{eq:degenerate_eigenvalues_largeNc}. As the construction is straightforward, we do not show the explicit expressions here.

With~\eqref{eq:resummed_odd_general} and~\eqref{eq:resummed_even_general}, we have extended the large-$N_c$ resummation of the Glauber series to processes with gluons in the initial state. It is remarkable that the resummed series can be expressed through simple trigonometric functions and at most fourfold integrals. These two equations, together with~\eqref{eq:resummed_odd_qq} and~\eqref{eq:resummed_even_qq}, are the main results of this work. As in the previous section, one may now proceed to determine the large-$w$ asymptotics by restricting oneself to the fixed-coupling approximation~\eqref{eq:fixed_coupling_approx}. Here, we omit a detailed discussion due to the complexity of the resulting expressions, which involve four-fold integrals that are no longer amenable to straightforward analytical evaluation. The qualitative dependence on the parameter $w$, however, is similar to~\eqref{eq:resummed_fixed_coupling_qq_odd} and~(\ref{eq:asymptotics_qq_even_1}--\ref{eq:asymptotics_qq_even_3}), i.e.~the leading term is constant up to logarithmic corrections.

\section{Numerical analysis}
\label{sec:numerics}

\begin{figure}[t]
\centering
\includegraphics[scale=1]{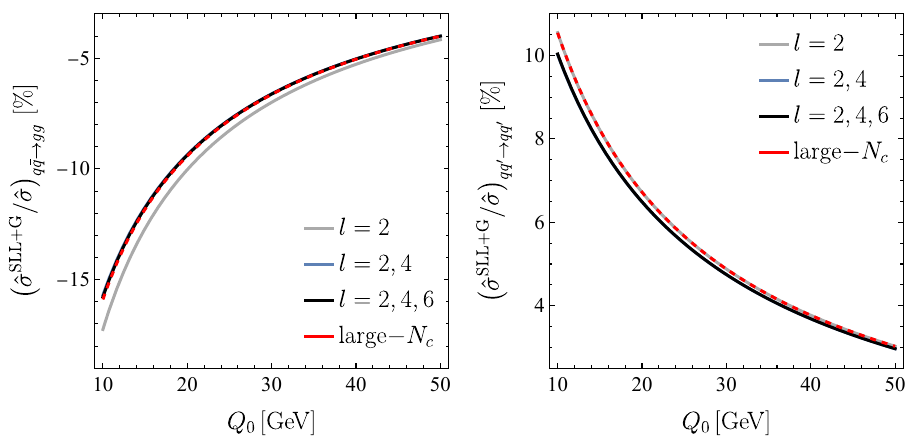}
\vspace{-1mm}
\caption{Numerical results for the Glauber-series contributions to partonic $q\bar q\to gg$ small-angle (left) and $qq'\to qq'$ forward scattering (right) as a function of the veto scale $Q_0$. Both plots show the SLL results (gray lines), the summed results for $l\leq4$ (blue lines), $l\leq6$ (black lines) and the result resummed in the large-$N_c$ limit~\eqref{eq:resummed_even_qq} (red lines). The blue and black curves overlap.}
\label{fig:LNc_qq_even}
\end{figure}

\begin{figure}[t]
\centering
\includegraphics[scale=1]{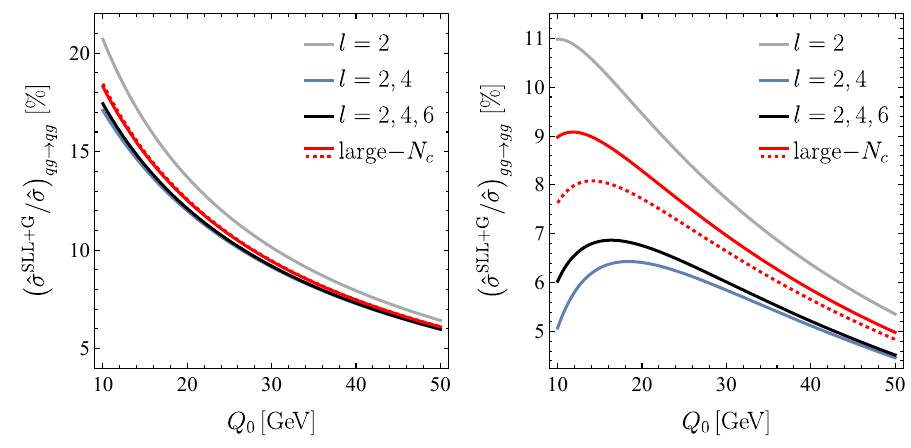}
\vspace{-1mm}
\caption{Numerical results for the Glauber-series contributions to partonic $qg\to qg$ forward (left) and $gg\to gg$ small-angle scattering (right) as a function of the veto scale $Q_0$. Both plots show the SLL results (gray lines), the summed results for $l\leq4$ (blue lines), $l\leq6$ (black lines) and the in the large-$N_c$ limit resummed result~\eqref{eq:resummed_even_general} (red lines). The red dotted lines show the resummed result in the large-$N_c$ limit but with full $N_c$ dependence of the SLLs.}
\label{fig:LNc_gluons_even}
\end{figure}

In Figures~\ref{fig:LNc_qq_even} and~\ref{fig:LNc_gluons_even}, we show the contribution of the resummed Glauber series to the partonic cross section $\hat{\sigma}_{2\to M}^{\rm SLL+G}$ defined in~\eqref{eq:sigmaSLLGpartonic}, normalized to the Born cross section $\hat\sigma$, for a few representative $2\to 2$ scattering processes. In each plot, the red line shows the result for the resummed Glauber series obtained in the limit of large $N_c$, derived in this work, whereas the black line corresponds to the numerical approximation of the analogous results with full $N_c$ dependence. The gray line gives the contributions of the SLLs only ($l=2$). As in our previous work~\cite{Boer:2024hzh}, we use a rapidity gap of width $\Delta Y=2$ and fix the hard and soft matching scales at $\mu_h=Q=1$\,TeV and $\mu_s=Q_0$, respectively. 

For practical reasons, the black line in each plot is obtained by summing up the contributions with $l=2,4$ and 6 Glauber phases. Yet higher-order terms are difficult to obtain numerically (as they involve eightfold and more integrations), but they are almost always negligible on the scale of the plots. To illustrate this fact, we show as a blue line the sum of the terms with $l=2$ and 4 Glauber phases. For quark-induced scattering processes, depicted in Figure~\ref{fig:LNc_qq_even}, the blue lines are almost invisible, as they are covered by the black ones. In scattering processes with gluons in the initial state, as shown in Figure~\ref{fig:LNc_gluons_even}, the contribution of the $l=6$ term is not necessarily negligible. The result for the sum of the infinite Glauber series including the full $N_c$ dependence would lie between the black and blue curves (closer to the black line). 

In Figure~\ref{fig:LNc_qq_even}, we focus on $q\bar q\to gg$ small-angle scattering (left panel) and $qq'\to qq'$ forward scattering (right panel). In the first case, the resummed result in the limit of large $N_c$ (red) is in almost perfect agreement with the result obtained including the full $N_c$ dependence (black). In the case of quark-quark scattering, on the other hand, the higher-order Glauber terms are absent in the limit of large $N_c$, and therefore the red line agrees with the $l=2$ term shown in gray.\footnote{Remember that we did not replace $\frac{2\spac C_F}{N_c} \to 1$ in~\eqref{eq:resummed_even_qq}, as would be demanded in the strict large-$N_c$ limit, thus capturing the contribution of the SLLs completely.} 
The difference between the black and red lines corresponds to a sizable $1/N_c^2$ correction to the leading term, which results from the (1,3) entry in the expression for the matrix $\mathbbm{V}^G$ in~\eqref{eq:GGammacVG}, which equals 4/9 for $N_c=3$.

Figure~\ref{fig:LNc_gluons_even} shows two examples of processes involving gluons in the initial state. Recall that in this case our treatment employs a strict large-$N_c$ expansion, in which all terms in the Glauber series (also the one with $l=2$) are subject to higher-order $1/N_c^2$ corrections. For forward quark-gluon scattering (left panel) the resummation of the Glauber series in the limit of large $N_c$ produces a result that comes very close to the exact one. The example of $gg\to gg$ small-angle scattering (right panel) is a bit pathological, since in this case high-order Glauber terms with $l=4$ and~6 still produce sizable contributions (see the blue and black lines), which are only captured partially in our new result derived in the limit of large $N_c$ (red line). For small values of $Q_0$, the red line overestimates the black one by approximately 50\%. The result can be improved by including the subleading terms in the $1/N_c$ expansion for the SLLs ($l=2$) but not for the higher-order Glauber contributions. The corresponding result is shown by the dotted red line, which reduces the discrepancy  to about 25\%.

\section{Conclusions}
\label{sec:conclusions}

Based on the formalism of~\cite{Boer:2024hzh}, and utilizing the basis of operators in color-space developed in~\cite{Boer:2023jsy,Boer:2023ljq}, we have presented closed-form expressions for the contributions of the resummed Glauber series to partonic scattering processes in renormalization-group improved perturbation theory, working at leading order in the limit of large $N_c$. As the SLLs and the Glauber series themselves are subleading in $1/N_c$ compared to e.g.~the non-global logarithms, the resummation performed here corresponds to a next-to-leading term in the large-$N_c$ expansion of the respective cross sections. For scattering processes initiated by quarks and/or anti-quarks, our results can be expressed as an at most two-fold integral over products of Sudakov-like evolution factors and basic trigonometric weight functions, as shown in~\eqref{eq:resummed_odd_qq} and~\eqref{eq:resummed_even_qq}. For processes featuring gluons in the initial state, the respective expressions~\eqref{eq:resummed_odd_general} and~\eqref{eq:resummed_even_general} are more complicated, containing up to three-fold (four-fold) integrations for quark-gluon-initiated (gluon-initiated) scattering processes. 

We have performed a numerical comparison of our resummed expressions obtained in the limit of large $N_c$ with (truncated) sums of the Glauber series with full $N_c$ dependence, finding that the analytic expressions derived here provide a good approximation of the full series in most cases and significantly improve the estimate of keeping only the super-leading logarithms. From a purely phenomenological point of view, however, a better approximation is obtained by including the contributions from the terms with the Glauber series with $l\le 4$, i.e.~by working with a truncated Glauber series, which involves as many integrations as the resummed large-$N_c$ result derived here.

More importantly, and from a conceptual point of view, the fact that the large-$N_c$ limit has allowed us to obtain simple closed-form expressions for the infinite Glauber series is a remarkable and unexpected result, since {\em a priori\/} each Glauber-operator insertion entails an additional scale integration. We have shown that almost all of them can be evaluated straightforwardly in the limit of large $N_c$. This finding suggests that large-$N_c$ methods might also be helpful to study other aspects of non-global logarithms at hadron colliders, such as secondary and yet higher-order soft emissions, which build up the series of non-global logarithms. Moreover, we are confident that the analytic expressions we have derived will be a crucial asset in the further development and validation of amplitude-level parton showers including quantum interference effects. 

\subsection*{Acknowledgments}

This research has received funding from the European Research Council (ERC) under the European Union’s Horizon 2022 Research and Innovation Program (ERC Advanced Grant agreement No.101097780, EFT4jets). Views and opinions expressed are however those of the authors only and do not necessarily reflect those of the European Union or the European Research Council Executive Agency. Neither the European Union nor the granting authority can be held responsible for them. The work reported here was also supported by the Cluster of Excellence \textit{Precision Physics, Fundamental Interactions, and Structure of Matter} (PRISMA$^+$, EXC 2118/1) within the German Excellence Strategy (Project-ID 390831469).

\clearpage
\begin{appendix}

\section{Quark-gluon-initiated processes}
\label{app:quark_gluon}

In this appendix, we list the relevant objects from the discussion in Section~\ref{sec:gluons} for quark-gluon-initiated processes and state explicit results for~\eqref{eq:resummed_odd_general} and~\eqref{eq:resummed_even_general}.

The matrix representations of the Glauber operator and the logarithmically-enhanced collinear anomalous dimension naturally decompose in the form~\cite{Boer:2023ljq} 
\begin{align} \label{eq:matrix_decomposition}
	\mathbbm{V}^G &=
	\begin{pmatrix}
		0 & \tilde{\nu}^{(j)} & 0 \\
		\nu^{(j)} & 0 & 0 \\
		0 & 0 & 0
	\end{pmatrix} ,
	&
	\GGammac &=
	\begin{pmatrix}
		\tilde{\gamma}^{(j)} & 0 & 0 \\
		0 & \gamma^{(j)} & 0 \\
		0 & \lambda & \gamma
	\end{pmatrix} .
\end{align}
In the color basis for quark-gluon-initiated processes rescaled with appropriate factors of $N_c$, see Appendix~A of~\cite{Boer:2024hzh}, the leading terms of $\mathbbm{V}^G$ in the large-$N_c$ expansion are 
\begin{align}
	\nu^{(j)} &=
	\begin{pmatrix}
		-1 & 0 & 0 \\
		2 & 0 & 0 \\
		-\frac{1}{2} & \frac{1}{2} & 0 \\
		\frac{1}{2} & -\frac{1}{2} & 0 \\
		1 & 0 & -1 \\
		0 & 0 & -1 \\
		0 & -1 & 0 \\
		0 & 0 & -1
	\end{pmatrix} ,
	&
	\tilde{\nu}^{(j)} &=
	\begin{pmatrix}
		0 & 0 & -\frac{1}{2} & \frac{1}{2} & 0 & 0 & 0 & 0 \\
		0 & 0 & \frac{1}{2} & -\frac{1}{2} & 0 & 0 & 0 & 0 \\
		0 & 0 & 0 & -1 & -1 & 0 & 0 & 0
	\end{pmatrix} .
\end{align}
For $\GGammac$ the leading terms of the four submatrices are 
\begin{equation}
\begin{aligned}
	\tilde{\gamma}^{(j)} &=
	\begin{pmatrix}
		1 & 0 & 0 \\
		0 & 1 & 0 \\
		0 & 0 & \frac{3}{2}
	\end{pmatrix} ,
	&\qquad
	\gamma^{(j)} &=
	\begin{pmatrix}
		\frac{1}{2} & 0 & 0 & 0 & 0 & 0 & 0 & 0 \\
		0 & \frac{1}{2} & 0 & 0 & 0 & 0 & 0 & 0 \\
		0 & 0 & 1 & 0 & 0 & 0 & 0 & 0 \\
		0 & 0 & 0 & 1 & 0 & 0 & 0 & 0 \\
		0 & 0 & 0 & 0 & \frac{3}{2} & 0 & 0 & 0 \\
		0 & 0 & 0 & 0 & -1 & \frac{3}{2} & 0 & 0 \\
		0 & 0 & 0 & 0 & 0 & 0 & \frac{1}{2} & 0 \\
		0 & 0 & 0 & 0 & 0 & 0 & 0 & \frac{3}{2}
	\end{pmatrix} ,
	\\
	\gamma &=
	\begin{pmatrix}
		0 & 0 & 0 \\
		0 & 1 & 0 \\
		0 & 0 & 1
	\end{pmatrix} ,
	&
	\lambda &=
	\begin{pmatrix}
		\frac{1}{2} & -\frac{1}{4} & 0 & 0 & 0 & 0 & 0 & 0 \\
		\frac{1}{2} & -\frac{1}{2} & 0 & 0 & -2 & \frac{1}{2} & 0 & 0 \\
		0 & 0 & 0 & 0 & 0 & 0 & \frac{1}{2} & -\frac{1}{2}
	\end{pmatrix} .
\end{aligned}
\end{equation}
After carrying out the matrix products in~\eqref{eq:coef_gluons}, one finds for odd $l$ 
\begin{equation}
\begin{aligned}
    \varsigma_1 &= \frac12 \, (1, -1, -1, \dots)^T ,
    \\
    \varsigma_{3/2} &= \frac12 \, (0, 0, -1, \dots)^T ,
\end{aligned}
\end{equation}
where we only show the first three components relevant for an odd number of Glauber-operator insertions, and for even $l$ 
\begin{equation}
\begin{aligned}
    \tilde\varsigma &= \frac12 \, (\dots, -2, 4, -1, 1, 2, 0, 0, 0, 0, 0, 0)^T ,
    \\
    \tilde\varsigma_1 &= \frac12 \, (\dots, -1, 2, -1, 1, 1, 0, 1, 0, 0, 0, 0)^T ,
    \\
    \tilde\varsigma_{3/2} &= \frac12 \, (\dots, 0, 0, 0, 0, 1, 1, 0, 1, 0, 0, 0)^T .
\end{aligned}
\end{equation}
Here we have dropped the first three components as they are irrelevant.
Since the eigenvalue $2$ does not appear for quark-gluon-initiated processes, i.e.\ $\mathbbm{V}^G_2=0$, one finds $\varsigma_2=\tilde\varsigma_2=0$.

Performing the products $\mathbbm{U}_c(\mu_h,\mu_1) \, \varsigma_{\spac i}$ in~\eqref{eq:resummed_odd_general}, the dependence on $\mu_1$ cancels for some terms and one can combine them by evaluating the $x_1$-integral with terms with one integral less.
The simplified result reads 
\begin{equation} \label{eq:resummed_odd_qg}
\begin{aligned}
    \sum_{\ell=1}^\infty \mathbbm{U}_{\rm SLL}^{(2\ell-1)}(\mu_h,\mu_s)\,\varsigma &= \frac{16i\pi}{\beta_0^2} \int_1^{x_s} \!\frac{dx_2}{x_2} \,\ln\frac{x_s}{x_2} \, \Bigg\{ U_c(1;\mu_h,\mu_2)
    \\[-2mm]
    &\qquad \times \bigg[\cos^2\!\bigg(\frac{\pi N_c}{\beta_0} \, \ln x_2\bigg)
    \begin{pmatrix}
        1 \\ 0 \\ 0 \\ \vdots
    \end{pmatrix}
    + \sin^2\!\bigg(\frac{\pi N_c}{\beta_0} \, \ln x_2\bigg)
    \begin{pmatrix}
        0 \\ 1 \\ 0 \\ \vdots
    \end{pmatrix} \bigg]
    \\
    &\quad + \frac{\pi N_c}{\beta_0} \int_1^{x_2} \!\frac{dx_1}{x_1} \, U_c(\textstyle{\frac32},1;\mu_h,\mu_1,\mu_2)
    \\[-2mm]
    &\qquad \times \bigg[\sin\!\bigg(\frac{2\pi N_c}{\beta_0} \, \ln x_2\bigg) + \sin\!\bigg(\frac{2\pi N_c}{\beta_0} \, \ln x_1\bigg) \bigg]
    \begin{pmatrix}
        0 \\ 0 \\ 1 \\ \vdots
    \end{pmatrix}
    \Bigg\} \,.
\end{aligned}
\end{equation}
In contrast to the result~\eqref{eq:resummed_even_qq} valid for quark-initiated processes, here we work in the strict large-$N_c$ limit.
This proves to be convenient, as there are no terms containing a threefold integral.

\section{Gluon-initiated processes}
\label{app:gluon}

In this appendix, we list the relevant objects from the discussion in Section~\ref{sec:gluons} for gluon-initiated processes and state explicit results for~\eqref{eq:resummed_odd_general} and~\eqref{eq:resummed_even_general}.

In the color basis for gluon-initiated processes rescaled with appropriate factors of $N_c$, see Appendix~A of~\cite{Boer:2024hzh}, the submatrices of $\mathbbm{V}^G$ in~\eqref{eq:matrix_decomposition} in the large-$N_c$ limit are given by 
\begin{align}
	\nu^{(j)} &=
	\begin{pmatrix}
		2 & 0 & 0 & 0 & 0 & 0 & 0 \\
		\frac{1}{2} & -\frac{1}{2} & 0 & 0 & 0 & 0 & 0 \\
		1 & 0 & -1 & 0 & 0 & 0 & 0 \\
		0 & 0 & -1 & -1 & 0 & 1 & 0 \\
		0 & 0 & 0 & 0 & 1 & 0 & -1 \\
		0 & 0 & 0 & \frac{1}{2} & 0 & \frac{1}{2} & 0 \\
		0 & 0 & 1 & -1 & 1 & 1 & -1
	\end{pmatrix} ,
	&
	\tilde{\nu}^{(j)} &=
	\begin{pmatrix}
		0 & 1 & 0 & 0 & 0 & 0 & 0 \\
		0 & -1 & 0 & 0 & 0 & 0 & 0 \\
		0 & -1 & -1 & 0 & 0 & 0 & 0 \\
		0 & 0 & 0 & 0 & 0 & 1 & 0 \\
		0 & 0 & -1 & 0 & 0 & 0 & \frac{1}{2} \\
		0 & 0 & 0 & 0 & 0 & 1 & 0 \\
		0 & 0 & 0 & 0 & 0 & 0 & -\frac{1}{2}
	\end{pmatrix} .
\end{align}
The ones of $\GGammac$ in this limit are 
\begin{equation}
\begin{aligned}
	\tilde{\gamma}^{(j)} &=
	\begin{pmatrix}
		1 & 0 & 0 & 0 & 0 & 0 & 0 \\
		0 & 1 & 0 & 0 & 0 & 0 & 0 \\
		0 & 0 & \frac{3}{2} & 0 & 0 & 0 & 0 \\
		0 & 0 & 0 & 2 & 0 & 0 & 0 \\
		0 & 0 & 0 & -1 & 2 & 0 & 0 \\
		0 & 0 & 0 & 0 & 0 & 2 & 0 \\
		0 & 0 & 0 & 0 & 0 & 0 & 2
	\end{pmatrix} ,
	&\quad
	\gamma^{(j)} &=
	\begin{pmatrix}
		\frac{1}{2} & 0 & 0 & 0 & 0 & 0 & 0 \\
		0 & 1 & 0 & 0 & 0 & 0 & 0 \\
		0 & 0 & \frac{3}{2} & 0 & 0 & 0 & 0 \\
		0 & 0 & -1 & \frac{3}{2} & 0 & 1 & 0 \\
		0 & 0 & 0 & 0 & \frac{3}{2} & 0 & \frac{1}{2} \\
		0 & 0 & 0 & 0 & 0 & 2 & 0 \\
		0 & 0 & 0 & 0 & 0 & -1 & 2
	\end{pmatrix} ,
	\\
	\gamma &=
	\begin{pmatrix}
		0 & 0 & 0 & 0 & 0 & 0 \\
		0 & 1 & 0 & 0 & 0 & 0 \\
		0 & 0 & 1 & 0 & -4 & 0 \\
		0 & 0 & 0 & 2 & -4 & 0 \\
		0 & 0 & 0 & 0 & 2 & 0 \\
		0 & 0 & 0 & 0 & 0 & 2
	\end{pmatrix} ,
	&
	\lambda &=
	\begin{pmatrix}
		-1 & 0 & 0 & 0 & 0 & 0 & 0 \\
		-1 & 0 & -4 & 1 & 0 & -2 & 0 \\
		0 & 0 & -1 & 0 & -1 & 0 & 1 \\
		0 & 0 & -1 & \frac{1}{2} & 0 & 0 & 1 \\
		0 & 0 & \frac{1}{4} & 0 & 0 & \frac{1}{2} & 0 \\
		0 & 0 & 0 & 0 & \frac{1}{2} & 0 & -\frac{3}{2}
	\end{pmatrix} .
\end{aligned}
\end{equation}
The relevant vectors created by applying powers of $\mathbbm{V}^G_i$ to $\varsigma$ in~\eqref{eq:coef_gluons} are for odd $l$ 
\begin{equation}
\begin{aligned}
    \varsigma_1 &= \frac12 \, (1, -1, -1, 0, 0, 0, 0, \dots)^T \,,
    \\
    \varsigma_{3/2} &= \frac12 \, (0, 0, -1, 0, -1, 0, 0, \dots)^T \,,
    \\
    \varsigma_2 &= \frac14 \, (0, 0, 0, 0, -1, 0, 1, \dots)^T \,,
\end{aligned}
\end{equation}
where we show only the first seven components relevant in this case.
For even $l$, the relevant components 8 to 14 of these vectors are 
\begin{equation}
\begin{aligned}
    \tilde\varsigma &= \frac12 \, (\dots, 4, 1, 2, 0, 0, 0, 0, \dots)^T \,,
    \\
    \tilde\varsigma_1 &= \frac12 \, (\dots, 2, 1, 1, 0, 0, 0, 0, \dots)^T \,,
    \\
    \tilde\varsigma_{3/2} &= \frac12 \, (\dots, 0, 0, 1, 1, 0, 0, -1, \dots)^T \,,
    \\
    \tilde\varsigma_2 &= \frac12 \, (\dots, 0, 0, 0, 0, -1, 0, -1, \dots)^T \,.
\end{aligned}
\end{equation}

Performing the products $\mathbbm{U}_c(\mu_h,\mu_1) \, \varsigma_{\spac i}$ in~\eqref{eq:resummed_odd_general}, the dependence on $\mu_1$ cancels for some terms and one can combine them by evaluating the $x_1$-integral with terms with one integral less.
The simplified result reads 
\begingroup
\makeatletter
\def\tagform@#1{\maketag@@@{\normalsize(#1)}}
\small
\begin{align}\label{eq:resummed_odd_gg_simplified}
    \sum_{\ell=1}^\infty \mathbbm{U}_{\rm SLL}^{(2\ell-1)}(\mu_h,\mu_s)\,\varsigma &= \frac{16i\pi}{\beta_0^2} \int_1^{x_s} \!\frac{dx_3}{x_3} \,\ln\frac{x_s}{x_3} \, \Bigg\{ U_c(1;\mu_h,\mu_3)
    \\\nonumber
    &\hspace{7mm} \times \bigg[\cos^2\!\bigg(\frac{\pi N_c}{\beta_0} \, \ln x_3\bigg) (1,0,0,0,0,0,0,\dots)^T
    \\\nonumber
    &\hspace{10mm} + \sin^2\!\bigg(\frac{\pi N_c}{\beta_0} \, \ln x_3\bigg) (0,1,0,0,0,0,0,\dots)^T
    \bigg]
    \\\nonumber
    &\quad + \frac{\pi N_c}{\beta_0} \int_1^{x_3} \!\frac{dx_1}{x_1} \, U_c(\textstyle{\frac32},1;\mu_h,\mu_1,\mu_3)
    \\\nonumber
    &\hspace{7mm} \times \bigg[\sin\!\bigg(\frac{2\pi N_c}{\beta_0} \, \ln x_3\bigg) + \sin\!\bigg(\frac{2\pi N_c}{\beta_0} \, \ln x_1\bigg) \bigg] (0,0,1,0,0,0,0,\dots)^T
    \\\nonumber
    &\quad + \bigg(\frac{\pi N_c}{\beta_0}\bigg)^{\!2} \int_1^{x_3} \!\frac{dx_2}{x_2} \int_1^{x_2} \!\frac{dx_1}{x_1} \, U_c(\textstyle{2,\frac32},1;\mu_h,\mu_1,\mu_2,\mu_3)
    \\\nonumber
    &\hspace{7mm} \times \Bigg( \bigg[\cos\!\bigg(\frac{2\pi N_c}{\beta_0} \, \ln\frac{x_3}{x_1}\bigg) + \cos\!\bigg(\frac{2\pi N_c}{\beta_0} \, \ln\frac{x_2}{x_1}\bigg) \bigg] (0,0,0,0,1,0,1,\dots)^T
    \\\nonumber
    &\hspace{10mm} + \bigg[\cos\!\bigg(\frac{2\pi N_c}{\beta_0} \, \ln x_3\bigg) + \cos\!\bigg(\frac{2\pi N_c}{\beta_0} \, \ln x_2\bigg) \bigg] (0,0,0,0,1,0,-1,\dots)^T
    \Bigg) \Bigg\} .
\end{align}
\endgroup%
Similar to~\eqref{eq:resummed_odd_qg} we work in the strict large-$N_c$ limit, as there are no terms containing a fourfold integral.

\end{appendix}

\clearpage
\pdfbookmark[1]{References}{Refs}
\bibliography{refs.bib}

\end{document}